\documentstyle[epsfig,prb,aps,twocolumn]{revtex}
\begin{document}
\tightenlines \wideabs{

\title{Superconductivity in the Chalcogens up to Multimegabar Pressures}
\author{Eugene Gregoryanz$^1$, Viktor V. Struzhkin$^1$, Russell J. Hemley$^1$,
Mikhail I. Eremets$^1$, \\ Ho-kwang Mao$^1$ and Yuri A.
Timofeev$^2$}
\address{$^1$Geophysical Laboratory and Center for High Pressure
Research, Carnegie Institution of Washington, \\
5251 Broad Branch Road NW, Washington D.C. 20015 U.S.A \\
$^2$Institute for High Pressure Physics, Russian Academy of Science, \\
142092 Troitsk, Moscow Region, Russia}
\maketitle

\begin{abstract}
Highly sensitive magnetic susceptibility techniques were used to
measure the superconducting transition temperatures in S up to
231($\pm$5) GPa. S transforms to a superconductor with T$_c$ of 10 K 
and has a discontinuity in T$_c$  dependence at 160 GPa
corresponding to bco to $\beta$-Po phase transition. Above this
pressure T$_c$ in S has a maximum reaching about 17.3($\pm$0.5) K
at 200 GPa and then slowly decreases with pressure to 15 K at 230 GPa.
 This trend in the pressure dependence parallels the behavior
of the heavier members Se and Te. Superconductivity in Se was also
observed from 15 to 25 GPa with T$_c$ changing from 4 to 6 K and
above 150 GPa with T$_c$ of 8 K. Similiarities in the T$_c$
dependences for S, Se, and Te, and the implications for oxygen are
discussed.
\end{abstract}
}

\narrowtext

\section{Introduction}

Comparative study of the high-pressure behavior of groups of
elements (i.e. with identical valences) provides detailed insight
into the effects of compression on electronic structure and its
control of physical properties. The chalcogen elements, or members
of group VIa, exhibit a variety of phases as a function of
pressures and temperatures and exhibit a broad range of
interesting physical properties.\cite{young} The lighter members
of the family form molecular crystals. The first is oxygen, which
forms a diatomic molecular crystal and persists in a variety of
molecular phases to at least 100 GPa. \cite{bunch} It also has
peculiar magnetic interactions, being the only magnetic insulator
among elements.\cite{young} At these higher pressures, however,
oxygen not only becomes metallic \cite{desgrenier} but is also
superconducting \cite{shimizu} above 95 GPa. Theoretical
calculations predict that the high-pressure $\zeta$-O$_2$ phase
above 96 GPa is a molecular metal, \cite{serra98} while
unambiguous experimental evidence of its molecular character is
still not available. Sulfur has one of the most complex diagrams
of the elements.\cite{young} Under ambient conditions, sulfur is
molecular and consists of S$_8$ species that form rings in the
crystalline state. Under pressure, sulfur undergoes a series of
both stable and metastable phase transitions up to $\sim$90~GPa,
finally becoming metallic at higher pressures.\cite{luo}

Under pressure the heavier group VIa elements (below oxygen)
undergo structural phase transitions involving non-molecular or
extended structures and become metallic (with the exception of Po
which is a metal at room pressure). Sulfur, selenium, and
tellurium follow similar trends in the sequence of observed
crystal structures. At comparatively low pressures, they
crystallize in base-centered orthorhombic (bco) structures, which
are metallic and superconducting. The situation is not as simple
at pressures below the stability range of the bco phase. The
sequence of phase transitions in Se below 30 GPa depends on the
starting phase of the material. \cite{akahama97} An unidentified
metallic phase was found in selenium (starting from pressures
about 13 GPa) which was found to be superconducting with T$_c$=5.2 K 
at 17 GPa.\cite{akahama97} Sulfur also has an unidentified phase
below 90 GPa, \cite{akahama93} which is however, semiconducting.
\cite{luo} On further compression in the bco phase, S, Se and Te
transform to the rhombohedral $\beta$-Po structure. If compressed
to even higher pressures at room temperature, Te and Se transform
to the body-centered (bcc) structure at 27 GPa (Ref.
\onlinecite{holz88}) and 140~GPa (Ref. \onlinecite{akahama93se}),
respectively.

The high-pressure structural and electronic properties of these
elements have been the subject of a variety of theoretical
studies. Pseudopotential total energy calculations \cite{zakharov}
have been performed for three high-pressure phases of S. They
suggested that S should transform from the $\beta$-Po to the bcc
phase at 545 GPa which would have a strong electron-phonon
coupling leading to a superconducting transition temperature of 15 K. 
More recent first-principles calculations \cite{rudin} predict
that compressed S favors the simple cubic structure over a wide
range of pressures from 280 GPa to 540 GPa before transforming to
a bcc phase. The later work also showed that upon entering the
simple cubic structure, the average phonon frequency $<\omega>$
increases substantially, leading to a smaller $\lambda$ and thus
smaller T$_c$. It was predicted that T$_c$ would drop below 10 K
upon entering the proposed simple cubic phase.\cite{rudin}

Recent experimental studies \cite{kometani,struzhkin} showed that
sulfur becomes metallic and superconducting at 93 GPa (with
T$_c$=10.1 K) and when pressurized above 160 GPa in $\beta$-Po
phase it has one of the highest known transition temperatures
among the elements (17 K).  Here we present a comparative study of
pressure-induced superconductivity in the chalcogens up to
multimegabar ($>$ 200 GPa) pressures. We extended the measurements
of superconductivity in S by magnetic susceptibility up to
231($\pm$5) GPa, which is the record pressure for such an
experiment. We also present results for T$_c$ in Se by direct
resistance measurements from 15 to 25 GPa and by magnetic
susceptibility up to 180($\pm$5) GPa in the bcc phase.

\section{Experimental Details}

\begin{figure}
\centerline{\epsfig{file=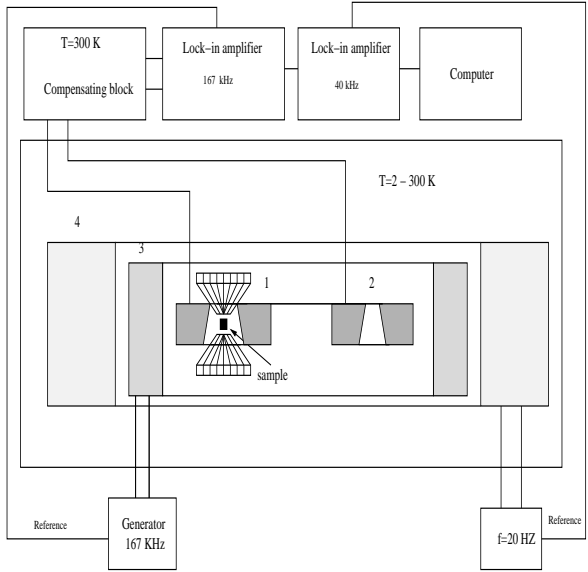,width=8cm,height=8cm}} \caption
{Schematic diagram of the diamond anvil cell magnetic
susceptibility technique.} 
\label{schematic}
\end{figure}

We performed the measurements of T$_c$ of S and Se using an
improved extension of a highly sensitive diamond-anvil cell
magnetic susceptibility technique described previously.
\cite{tim92} This technique is based on the quenching of
superconductivity and suppression of the Meisner effect in the
sample by an external magnetic field. The susceptibility of the
metallic diamagnetic parts of the diamond cell is essentially
independent of the external field. Applying a magnetic field to
the diamond cell (and sample) will therefore change the signal
coming from the sample while the background arising from the
surrounding diamagnetic parts will be nearly constant. The
measurements are done with four coils (shown in Fig.
\ref{schematic}). Two small coils consist of a signal coil, which
is wound around the sample (1) and a compensating coil (2)
connected in opposition. The excitation coil (3) encompasses both
the signal and compensating coils. The alternating high-frequency
magnetic field at the signal coil is created by the excitation
coil fed from the high-frequency generator. The alternating
magnetic field excites electromotive forces in the signal coil.
The fourth coil (4) placed around the diamond cell is used to
destroy the superconductivity in the sample near the
superconducting transition by application of a low-frequency ({\it
f}=20 Hz) magnetic field with an amplitude of several tens of
Oersteds. This leads to a change in magnetic susceptibility of the
sample from -1 to 0 twice in a given period and produces a
modulation of the signal amplitude in the signal coil with a
frequency 2{\it f}. The lock-in technique is used to record this
signal as a function of temperature. The technique sensitivity was
improved recently by using higher modulation frequency in the
excitation-pickup coil set up. \cite{timofeev}

Samples of 99.9995\% purity S were loaded in Mao-Bell cells
\cite{mao_cell} made from Be-Cu and modified for measurements down
to liquid helium temperatures. The gaskets made from nonmagnetic
Ni-Cr alloy were used together with tungsten inserts to confine
the sample and no pressure transmitting medium was used. The
gasket and insert may be responsible for the temperature dependent
background seen in the raw temperature scans (e.g., Fig.
\ref{raw}). To reach pressures above 200rGPa we used beveled
diamonds with 50 $\mu$m flats and 300 $\mu$m outer culets. The
initial sample size was $\sim$35 $\mu$m in diameter and $\sim$10 $\mu$m 
thickness. Pressures were measured by the ruby fluorescence
technique (quasihydrostatic scale) \cite{mao_ruby} using Ar$^+$
and Ti-sapphire laser excitation {\it in situ} at low
temperatures.

For Se, we used direct conductivity measurements of sample
resistance up to 35 GPa. \cite{eremets} Four electrical leads
formed from platinum foil allowed four-electrode measurements of
the resistance. To insulate the electrodes from the metallic
gasket, an insulating layer made from the mixture of cubic boron
nitride powder and epoxy was used. The superconducting transition
was detected by direct resistance measurements and by using a
modulating technique similar to that used in the susceptibility
experiments. At higher pressures (up to 180 GPa) we used the
magnetic susceptibility technique described above. The
configurations of the cell, diamonds, gasket and ruby fluorescence
imeasurement technique were identical to that used in the S
experiments.

\section{Results and Discussion}
\subsection{Superconductivity in S to 230 GPa}
\begin{figure}
\centerline{\epsfig{file=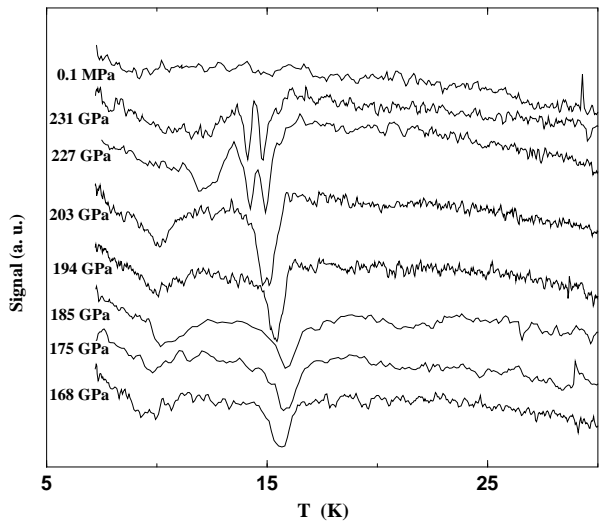,width=8cm}} \caption {Typical
signal recorded for S during low-temperature scans. The top scan
(background) was measured after the high-pressure run.}
\label{raw}
\end{figure}

\begin{figure}
\centerline{\epsfig{file=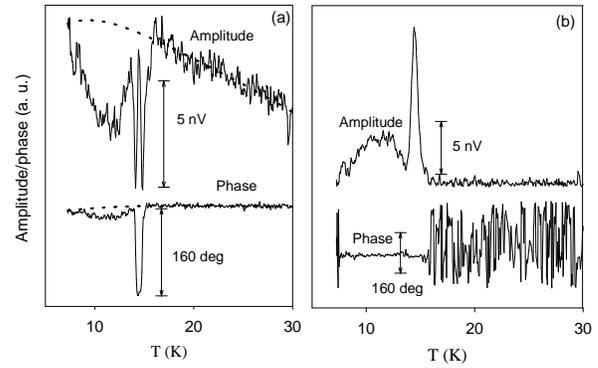,width=8cm}} \caption {Procedure
for background subtraction. (a) Total signal can be represented as
the complex variable ${\bf U} = A_{S+B}e^{i\phi_{S+B}}$ and the
interpolated background is just ${\bf B} = A{_B}e^{i\phi_B}$. (b)
The signal is equal to ${\bf S=U-B}$} 
\label{back}
\end{figure}

\begin{figure}
\centerline{\epsfig{file=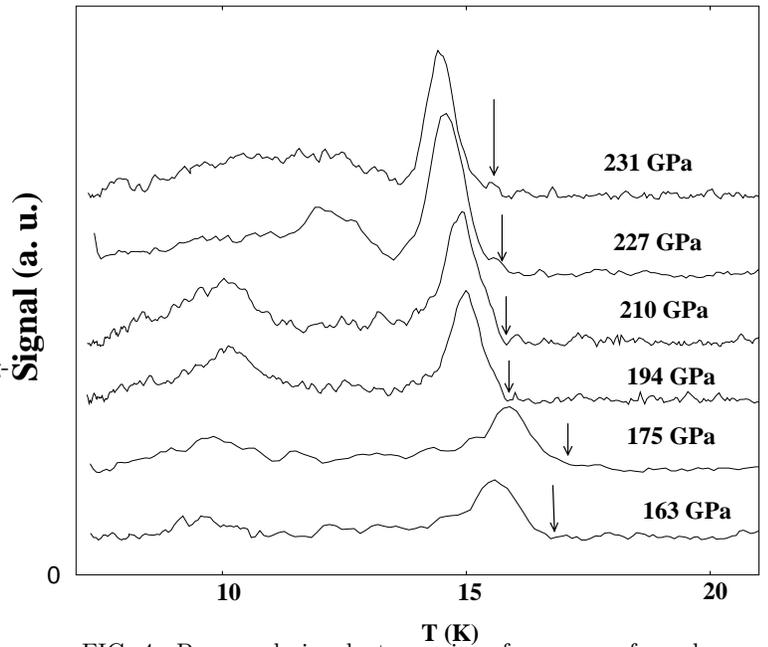,width=10cm}} \caption {Processed
signal at a series of pressures for sulfur. The wide peak at lower
temperatures is from the sample outside of the culet.}
\label{Ssignal}
\end{figure}

Several runs were made with decreasing sample sizes and diamond
culet dimensions to successively higher pressures. Measurements
are made as a function of temperature. Representative results are
shown in Figs. \ref{raw}-\ref{Ssignal}. No superconductivity was
detected up to approximately 90~GPa. At 93 GPa, however, a peak
characteristic of the supercondcting transition to the
sperconducting state was observed.\cite{tim92} T$_c$ is identified
as the temperature where the signal goes to zero on the high
temperature side (e.g., Fig.\ref{Ssignal}), which is the point at
which magnetic flux completely enters the sample. \cite{tim92} Two
peaks are clearly seen at $\sim$10-12 K and $\sim$17 K. The second
broad peak at lower temperatures arises from the sample outside of
the flat culet, where pressure is considerably lower than in the
middle of the culet. At pressures over 190 GPa the peak at 17 K
becomes visibly split. This splitting is artificial and only
reflects the fact that the signal amplitude has increased
substantially with respect to the background.

The background signal in our measurements appeared to be
ferromagnetic, as its phase is approximately opposite to that of
the signal from the sample (diamagnetic) (see Fig. \ref{raw}).
Because the background signal changes smoothly with temperature,
we can separate the signal from the backround by the simple
procedure illustrated in Fig. \ref{back}. We measure amplitude and
phase of a sum of signal and background with the lock-in
technique. The signal changes very abruptly in the vicinity of the
superconducting transition, allowing us to see these changes both
in amplitude and phase (Fig. \ref{back}). It is straightforward to
interpolate the background in the range of the superconducting
transition with a smooth polynomial function. The total signal can
be represented as the complex variable ${\bf U} =
A_{S}e^{i\phi_{S}}$, and the interpolated background as ${\bf B} =
A{_B}e^{i\phi_B}$; our signal is then ${\bf
S}=A_{S}e^{i\phi_S}$={\bf U-B} (the difference of two complex
variables).

The signal with background subtracted is shown in Fig.
\ref{Ssignal}. As in previous work, \cite{struzhkin} we observed
the appearance of a T$_c$ signal at $\sim$93 GPa at 10 K. T$_c$
gradually increases and upon entering the $\beta$-Po phase at 165~GPa 
it jumps to 17.0($\pm$0.5) K. For the next 35 GPa, it still
slowly increases, reaching the a weak maximum of 17.3($\pm$0.5) K
at 180-200 GPa; it then decreases on further compression, dropping
to 15 K at 231 GPa.

\subsection{Superconductivity in Se to 180~GPa}

\begin{figure}
\centerline{\epsfig{file=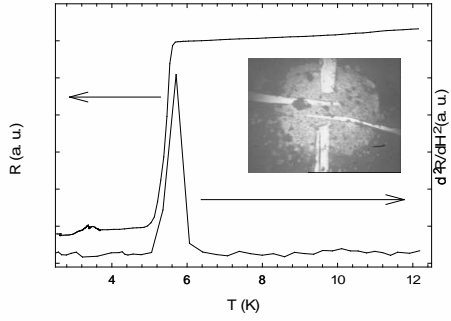,width=10cm}} \caption
{Temperature dependence of resistance and resistance modulated by
the magnetic field for Se. Inset: micrograph of the 4-probe
electrodes arrangement used in these measurements.}
\label{Sesignal}
\end{figure}

Two sets of experiments were performed on Se. In the first
experiment, we used 99.999\% pure amorphous Se as the starting
material with NaCl as pressure transmitting medium. Raman spectra
showed that Se crystallized at around 11 GPa.\cite{bandit} A
measurable T$_c$ signal appeared at $\sim$25 GPa at 5.8 K, in
agreement with previous studies.\cite{akahama92} Figure
\ref{Sesignal} shows the dependence of resistance and its
modulations by the magnetic field as a function of temperature. In
both cases, the onset of the superconducting transition in the
sample is clearly seen. After reaching 35 GPa, the pressure was
released. The T$_c$ signal was observable down to 18 GPa and
decreased with releasing pressure.

In the second set of experiments amorphous Se was pressurized up
to 180 GPa and T$_c$ was measured by susceptibility technique.
Currently, it is difficult to measure T$_c$ signals of $\sim$30 $\mu$m 
in diameter samples with this technique if the temperature
of superconducting transition is below 4 K due to a developing
background signal which is probably related to paramagnetic signal from 
the gasket (tungsten inset and Ni-Cr-Al alloy gasket material). 
We were able to detect the superconducting signal in the bcc phase 
from 140 to 180 GPa. On decompression below 130 GPa, the measurable 
signal disappeared.

\subsection{Comparison of T$_c$ among the chalcogens}

\begin{figure}
\centerline{\epsfig{file=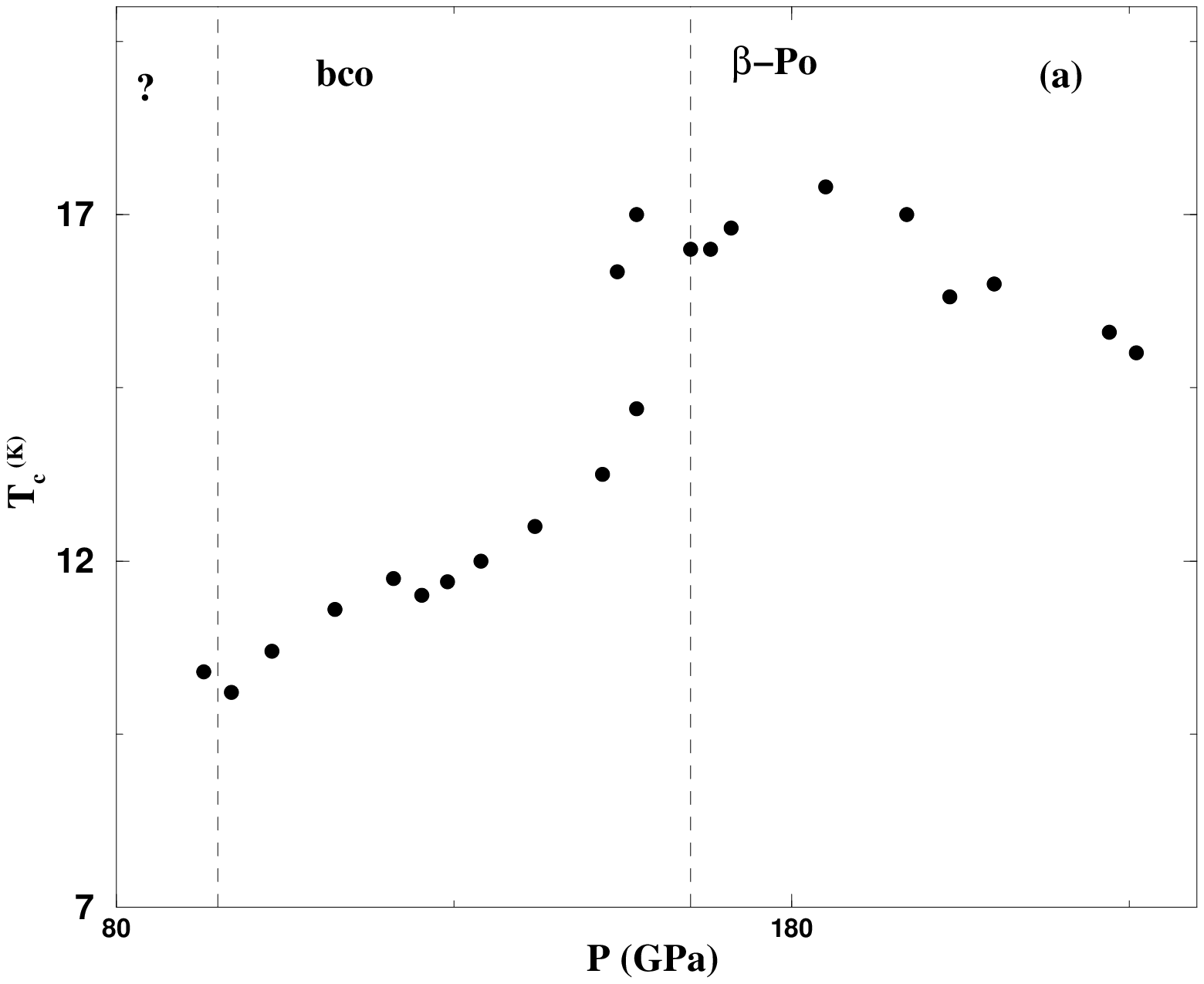,height=6cm,width=8cm}}
\centerline{\epsfig{file=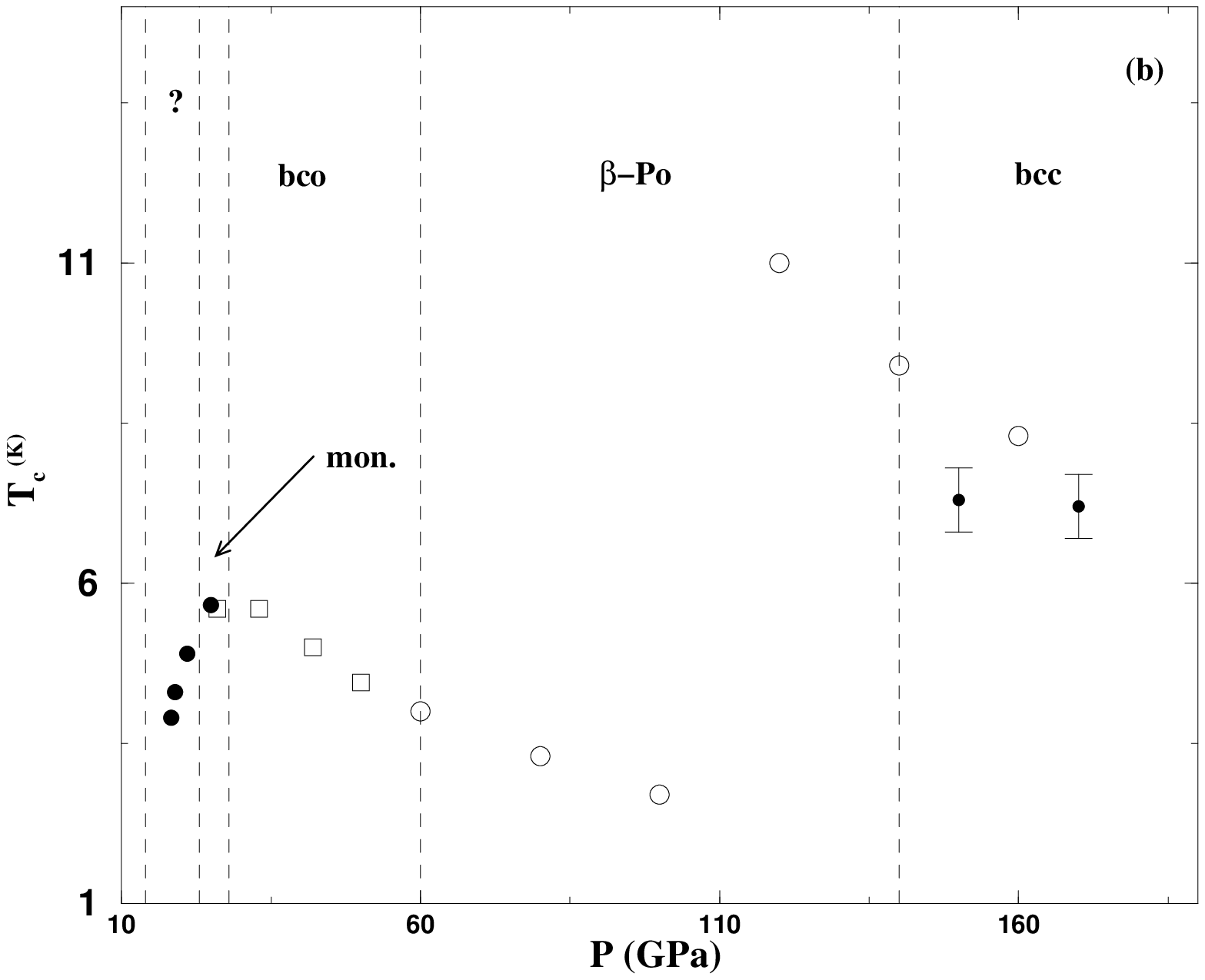,height=6cm,width=8cm}}
\centerline{\epsfig{file=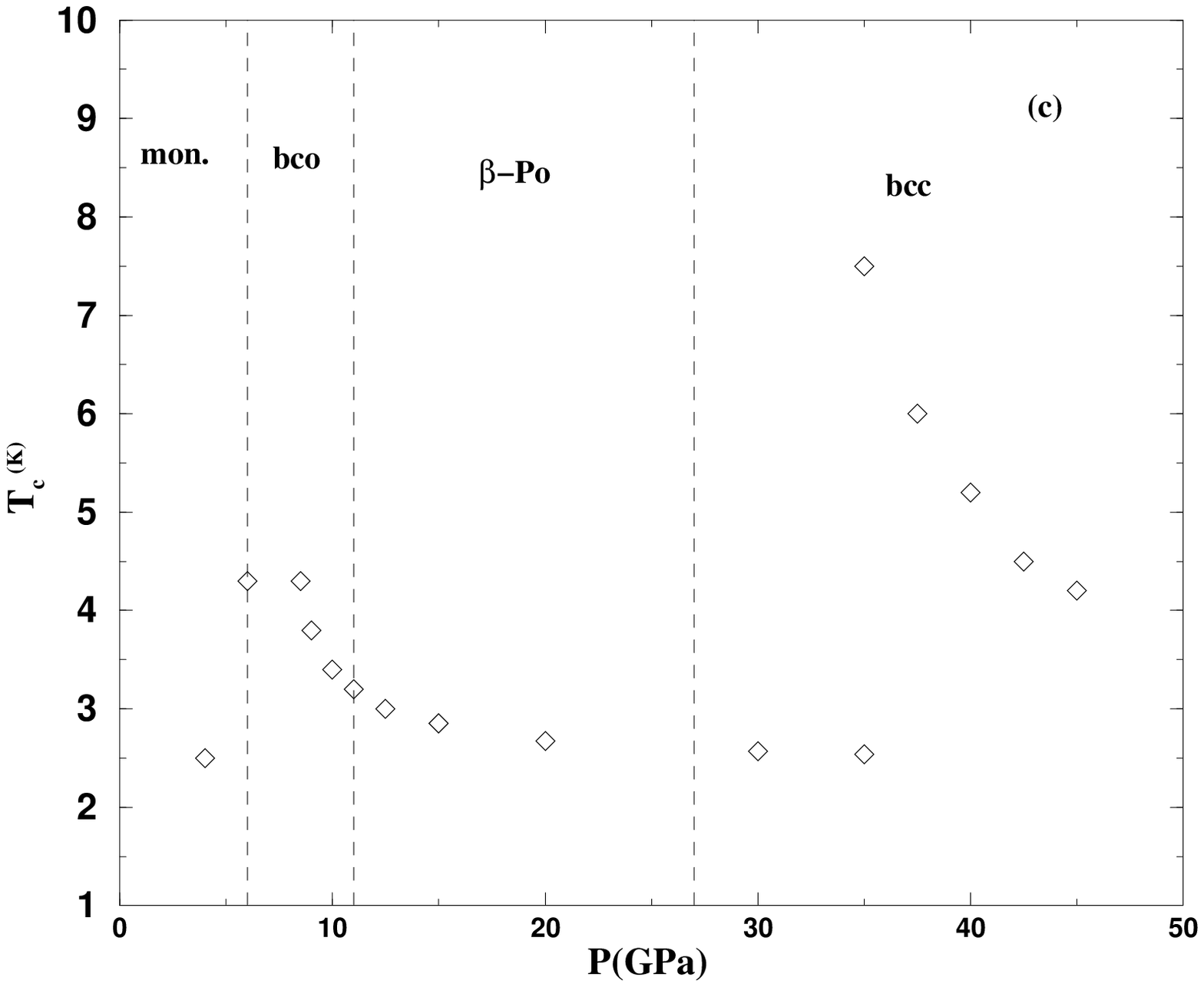,height=6cm,width=8cm}} \caption
{Temperature dependences of T$_c$(P) in (a) S, (b) Se, and (c) Te.
The experimental points for Se and Te shown as open diamonds are
from Refs. [\protect\onlinecite{akahama92,berman}]. The
theoretical calculation for Se shown in (b) with open circles are
from Ref. [\protect\onlinecite{amy}]. The dashed lines are the
phase boundaries measured at room temperature by x-ray
diffraction.} i
\label{TcvsP}
\end{figure}

\begin{figure}
\centerline{\epsfig{file=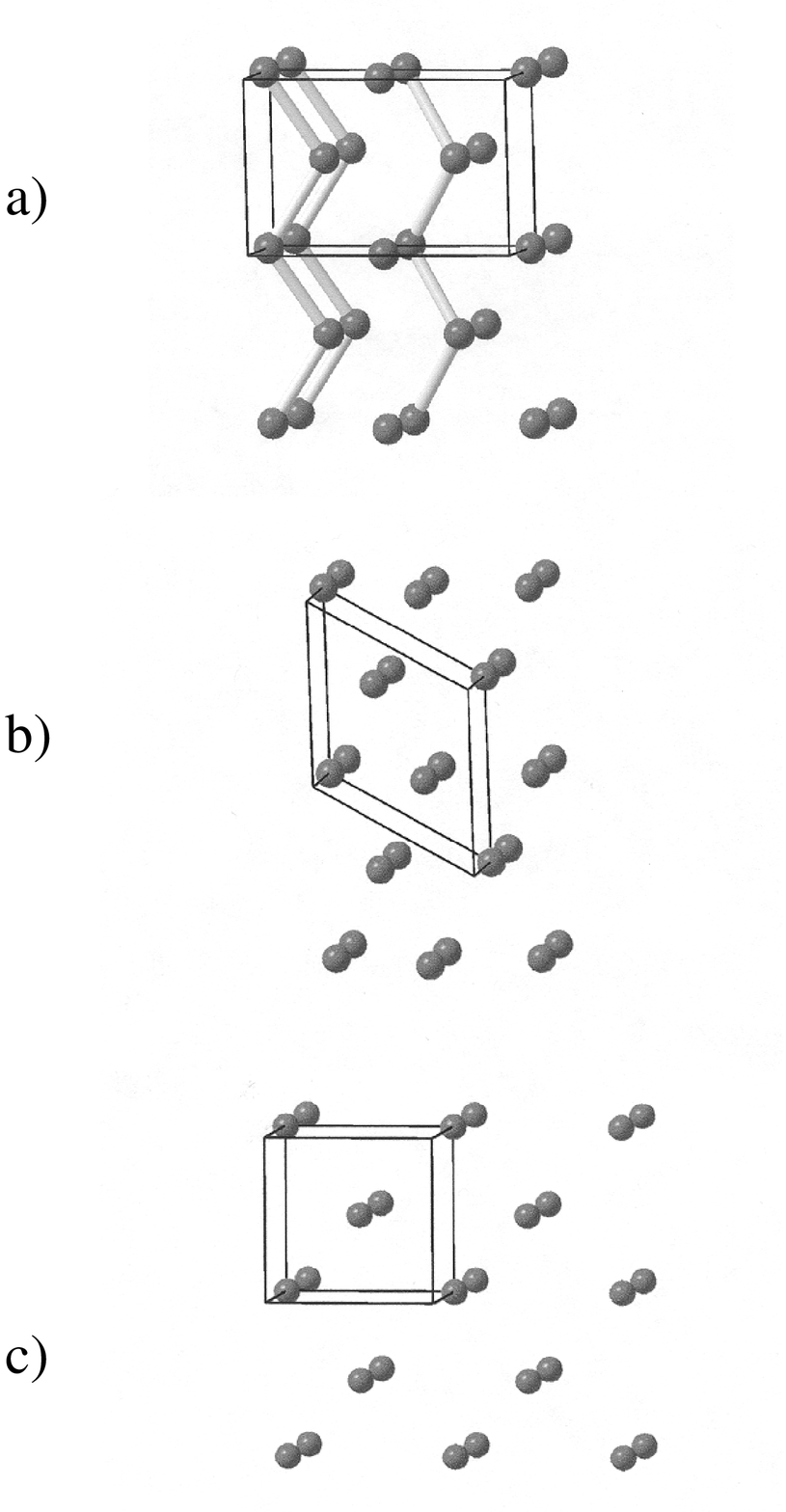}} \caption{Comparison between
the (a) bco, (b) $\beta$-Po, and (c) bcc structures.}
\label{struct}
\end{figure}

The pressure dependences of T$_c$ in the group VIa elements are
shown in Fig. \ref{TcvsP}, together with the observed transition
pressures between the phases. It should be noted that these
boundaries were determined by x-ray diffraction at room
temperature while the observations of superconductivity were done
at temperatures ranging from 2 to 18 K. It may be assumed that the
phase boundaries at low temperatures (if any) would be shifted in
pressure. The observed structures are illustrated in Fig.
\ref{struct}.

Tellurium becomes superconducting in the monoclinic phase
at $\sim$4 GPa. As shown in Fig. \ref{TcvsP}c, T$_c$ rises linearly with
pressure in this phase upon transforming to the bco structure, it
levels off and starts to decrease, passing through the field of
stability of the $\beta$-Po phase. T$_c$ then jumps by almost
factor of three and starts to come down again. This jump in T$_c$
happens at 35 GPa while room temperature phase transition to bcc
phase happens at 27 GPa.

Under ambient conditions, Se can be found in trigonal, monoclinic
or amorphous forms. Akahama {\it et al}. \cite{akahama97} showed
that monoclinic and trigonal Se become metallic at 12 and 23 GPa
respectively. Under pressure amorphous Se crystallizes in a
trigonal structure and becomes metallic at 12 GPa. It was shown
\cite{akahama97} that if $\alpha$-monoclinic Se used as the
starting material, it transforms at 12 GPa to an unidentified
metallic phase which is superconducting at 17 GPa but no pressure
dependence of T$_c$ was given. Figure \ref{TcvsP}b shows our
results for Se combined with the results of Ref. \cite{akahama92}
The trend in the T$_c$ dependence for Se is remarkably similar to
that of Te, although unlike the latter, Se becomes metallic and
superconducting in an unidentified phase from 14 to 23 GPa. The
structural sequence is identical to that of Te starting from 23 GPa 
where Se enters the monoclinic phase. Unlike the situation for
Te, T$_c$ within this phase changes little with pressure. When Se
is compressed further, the abrupt decrease in T$_c$ starts at 33 GPa 
which can be attributed to the transition to the bco phase,
which happens at 27 GPa at room temperature. There are no
experimental measurements of T$_c$ for Se from 60 to 150 GPa, but
theoretical calculations \cite{amy} predict a decrease of T$_c$ in
the $\beta$-Po phase, followed by an increase to much higher
values around 140 GPa where Se transforms to the bcc structure at
300 K. The values of T$_c$ in Se at 150-170 GPa measured here are
very close to those calculated in Ref. \onlinecite{amy}.

The structures of the semiconducting phases of S below the 90 GPa
transition is not known. It becomes metallic and superconducting
in the bco phase. Unlike the behavior of Se and Te, T$_c$ in this
phase of S increases with pressure and jumps abruptly upon
entering the $\beta$-Po-type phase (Fig. \ref{TcvsP}a). In the
$\beta$-Po phase, T$_c$ slowly decreases with further compression.
This decrease is consistent with the theoretical calculations of
Rudin {\it et al.} \cite{rudin}, which predicted a decrease in the
electron-phonon coupling parameter $\lambda$ as the sc phase is
approached.\cite{rudin}

It can be clearly seen from the behavior of the T$_c$ for the
three elements that the transition temperatures cannot be simply
explained by the existence of phase transitions and similar values
for electron-phonon coupling alone. On the other hand, the similar
trends are broadly consistent with changes towards higher symmetry
structures (transformation of layered phases to the close packed
structures at the highest pressures) with weakening of the
directional covalent bonding, as we now discuss.

M{\"o}ssbauer studies of Te showed a decrease in the quadrupole
splitting with pressure that was ascribed to
strengthening interactions between the neighboring chains and
weakening of the covalent bonds within the chain.\cite{vulliet}
The splitting disappeared when Te was pressurized into the bcc
phase where no covalent bonding is present. Thus we assume that 
the continuous weekening of the covalent bonds with increasing 
pressure is due to increasing screening by free carriers  
due to increasing density of states at the Fermi level.

\begin{figure}
\centerline{\epsfig{file=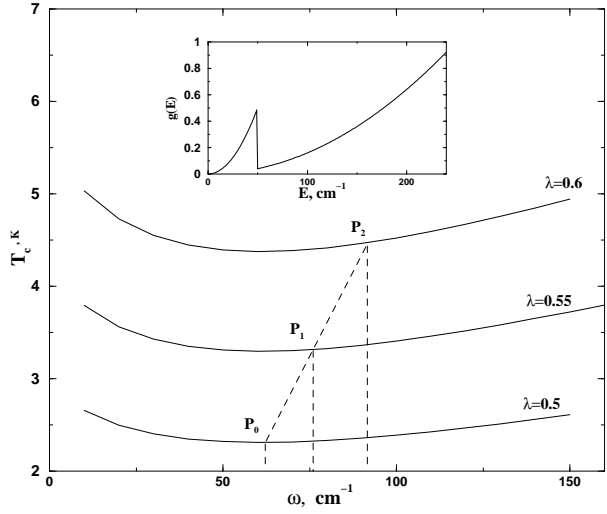,width=8cm}} \caption {T$_c$ as a
function of the pressure dependent Debye frequency. Inset: the
density of states used to calculate $<\omega_{log}>$. The dashed
line shows the increase in T$_c$ corresponding to the increase in
Debye frequency and $\lambda$, which was varied from 0.5 to 0.6.
The increase in frequency corresponds to the increase in pressure
by $\sim$20\%-30\%.} 
\label{Tcmodel}
\end{figure}

To analyse the possible T$_c$ increase in the lower pressure 
phases of the chalcogens due to the change in phonon spectra, 
we estimated T$_c$ in the lower symmetry "layered" 
structures (i.e monoclinic, bco) using the Allen-Dynes
formula \cite{allen} with literature values of $\lambda$
(electron-phonon coupling constant) and $\mu^{*}$ (effective
Coulomb repulsion potential). We assumed $\mu^{*}$=0.1 and varied
$\lambda$. The values of $<\omega_{log}>$ and $<\omega^2>$, the averages of the
phonon frequencies weighted to represent the strength of the
electron-phonon coupling, were estimated  with the phonon 
density of states shown in Fig. \ref{Tcmodel}. The density of states was
assumed to have  two cutoff Debye frequencies with values typical
for Se and Te corresponding to intralayer bond-stretching and inter-layer vibrations. 
In our estimates of T$_c$ versus pressure, the lower energy peak which we 
attributed to interlayer vibrations was allowed to move towards higher energies 
with pressure increase while the higher energy peak 
representing intralayer bond-stretching  modes was assumed 
to be pressure-independent.  The values of T$_c$ obtained in these
calculations are within the range of the experimentally observed
T$_c$ for Se and Te in their monoclinic and bco structures. However, T$_c$ is
virtually independent of phonon spectra but instead depends on $\lambda$.
From experimental data, $\lambda$ increases with pressure, meaning 
that $N(E_F)<I^2>$ increases faster  than $M<\omega^2>$ (which does not change 
substantially in our simplified model), where$\lambda=\frac{N(E_F)<I^2>}{M<\omega^2>}$,  
according to McMillan \cite{mcmil}. Here $N(E_F)<I^2>$ is the 
Hopfield parameter, which is generally assumed to be inversely 
proportional to volume in simple s-p metals \cite{mcmil}.
Thus, the increase in T$_c$ observed in the
monoclinic phase of Te and the unidentified phase of Se where
covalent bonding is still significant is most probably governed by the 
increase in Hopfield parameter with pressure. 
Sulfur has increasing T$_c$ in its bco phase where covalent bonding 
is still present.

The bco $\rightarrow$ $\beta$-Po $\rightarrow$ bcc structural 
sequence may also be possible for oxygen, although the stability 
of its diatomic state appers to extend to very high pressures
($>$ 100 GPa \cite{loubere}. It remains 
to be determined at what pressures the oxygen molecules dissociate (at low
temperature), and what crystal structures would form. The
additional complication with oxygen is its magnetic properties,
which could play important role in determining T$_c$. If oxygen
follows a similar structural trend in its nonmolecular
state, a similar T$_c$ dependence might be expected. 
Recent calculations \cite{otani} suggest that could be 
in  1 TPa range. 

\subsection{Conclusions}

We have measured T$_c$ of S up to 231 GPa, a record pressure for
both superconductivity and magnetic susceptibility techniques. The
value of T$_c$ at the highest pressures (15 K above 200 GPa, decreasing 
from 17 K at 160 GPa) in accord with the original theoretical 
predictions for the hypothetical bcc phase \cite{zakharov} and 
more recent calculations for the $\beta$-Po structure \cite{rudin}.
The superconducting behavior of Se was observed in an unidentified
phase from 14 to 20 GPa. The T$_c$ for Se is also
in very good agreement with recent first-principles 
calculations \cite{amy} (assuming $\beta$-Po, bcc, fcc structures).
The data obtained in the present work, previously published
experimental data, and recent theoretical calculations for the
chalcogens allow us to conclude that at lower pressures T$_c$ is
controlled at least in part by changes in covalent bonding. In 
higher pressures the bcc phases of these materials, 
the behavior of T$_c$ resembles normal {\it p} metals and 
decreases with pressure.

\subsection{Acknowledgements}
We thank A. Liu for sharing  results on Se prior to publication and
S. Rudin and P. Dera for helpful discussions. This work was
supported by the National Science Foundation.


\clearpage
\newpage

\end{document}